\documentclass[journal,onecolumn]{IEEEtran}

\usepackage[space]{cite}

\ifCLASSINFOpdf
   \usepackage[pdftex]{graphicx}
\else
 \usepackage[dvips]{graphicx}
\fi
\usepackage{graphicx}

\usepackage[cmex10]{amsmath}
\usepackage{txfonts}
\usepackage[mathscr]{euscript}
\usepackage{amsfonts}
\usepackage{bm}
\usepackage{longtable}
\usepackage{algorithmic}
\usepackage{algorithm2e}
\usepackage{setspace}
\usepackage{rotating}
\usepackage{amsfonts}
\usepackage{tabularx}
\usepackage{enumerate}
\usepackage[table]{xcolor}
 \usepackage{epsfig}
\usepackage{epstopdf}
\usepackage[T1]{fontenc}
\usepackage{mwe}    
\usepackage{xcolor}
\usepackage{color,soul}
\usepackage{yfonts}
\usepackage{xr}
\usepackage{url}
\usepackage[normalem]{ulem}
\usepackage{subfig}
\usepackage{float}
\usepackage{rotating}
\usepackage{pdflscape}
\let\oldbibliography\thebibliography
\renewcommand{\thebibliography}[1]{%
  \oldbibliography{#1}%
  \setlength{\itemsep}{-1pt}%
}
\makeatletter
\renewcommand{\@algocf@capt@plain}{above}
\renewcommand{\figurename}{Figure}
\makeatother
\input epsf

\begin{document}
\title{Bender's Decomposition for Optimization Design Problems in Communication Networks}
\author{Ahmed~Ibrahim, Octavia~A.~Dobre, Telex M. N. Ngatched, and~ Ana~García~Armada
\thanks{A. Ibrahim, O. A. Dobre and T. M. N. Ngatched are with the Faculty of Engineering and Applied Science, Memorial University of Newfoundland, St. John's, NL, Canada, e-mails: (amaibrahim@mun.ca, odobre@mun.ca ,tngatched@grenfell.mun.ca)}
\thanks{A.~G.~Armada is with the Department of Signal Theory and Communications,
Universidad Carlos III de Madrid, Avda. de la Universidad 30 (28911 Leganés) Spain  e-mail:ana.garcia@uc3m.es}}
\maketitle
\vspace{-1cm}
\begin{abstract}Various types of communication networks are constantly emerging to improve the connectivity services and facilitate the interconnection of various types of devices. This involves the development of several technologies, such as device-to-device communications, wireless sensor networks and vehicular communications. The various services provided have heterogeneous requirements on the quality metrics such as throughput, end-to-end latency and jitter. Furthermore, different network technologies have inherently heterogeneous restrictions on resources, e.g., power, interference management requirements, computational capabilities, etc. As a result, different network operations such as spectrum management, routing, power control and offloading need to be performed differently. Mathematical optimization techniques have always been at the heart of such design problems to formulate and propose computationally efficient solution algorithms. One of the existing powerful techniques of mathematical optimization is Benders decomposition (BD), which is the focus of this article. Here, we briefly review different BD variants that have been applied in various existing network types and different design problems. These main variants are the classical, the combinatorial, the multi-stage, and the generalized BD. We discuss compelling BD applications for various network types including heterogeneous cellular networks, infrastructure wired wide area networks, smart grids, wireless sensor networks, and wireless local area networks. Mainly, our goal is to assist the readers in refining the motivation, problem formulation, and methodology of this powerful optimization technique in the context of future networks. We also discuss the BD challenges and the prospective ways these can be addressed when applied to communication networks’ design problems. 
\end{abstract}

\section {Introduction}

Nowadays, communication networks have become highly complex due to the requirement to support heterogeneous applications. This impacts the decision-making encountered in various network operations such as routing, resource allocation, and user association. Specific design algorithms, which depend on the network type, its characteristics, and the user/application quality-of-service requirements are hence needed.

\begin{figure*}[ht]
	\begin{center}
	\centering
\includegraphics[width=1.0\textwidth]{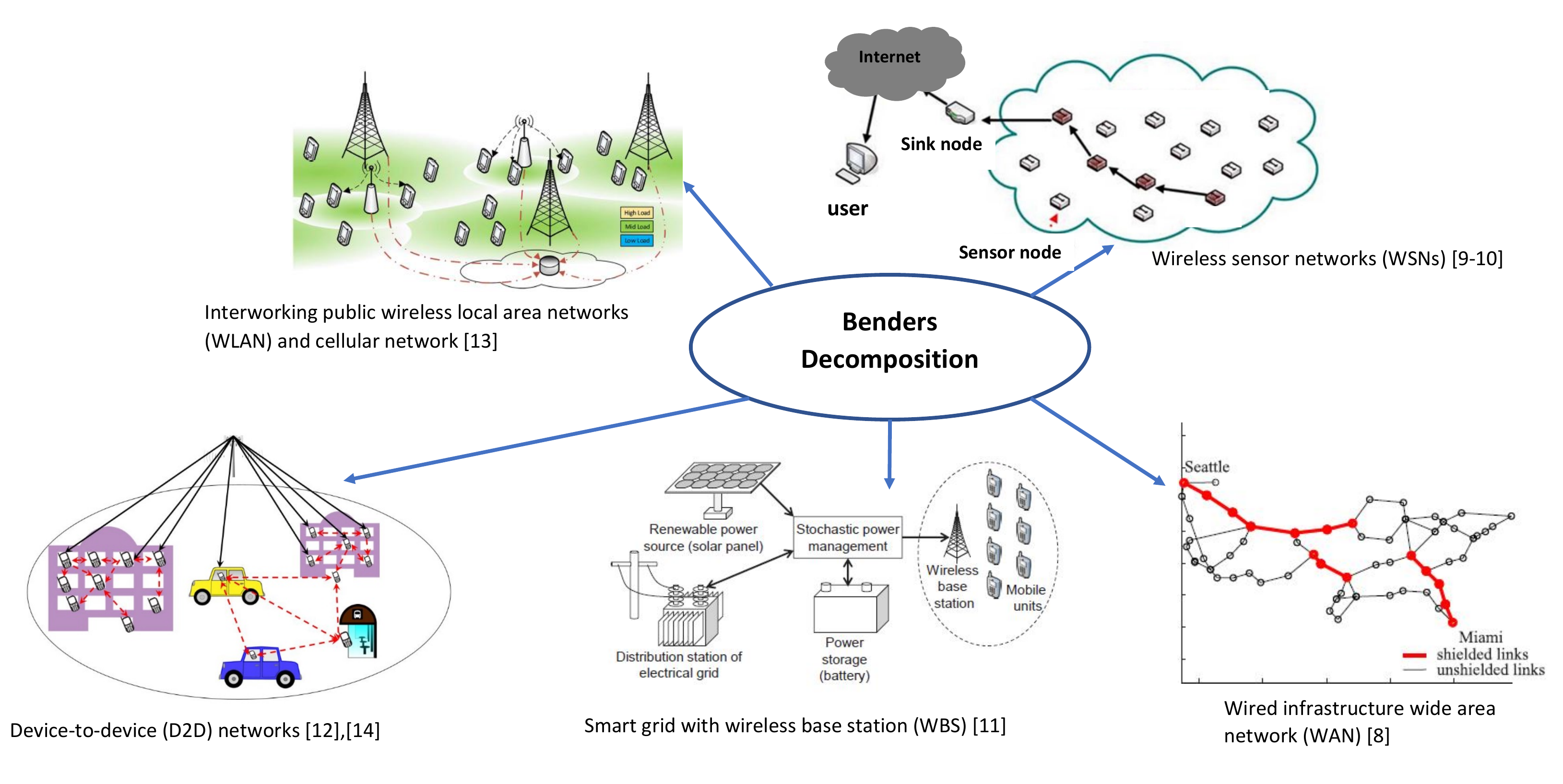}%
\caption{Different network types in which BD applications exist.}
\label{net_types}
\end{center}
\end{figure*}
\textit{Mathematical programming} techniques are widely used decision-making tools regardless of the network type. A decision is made such that a certain criterion is optimized (e.g., end-to-end delay) given some system related constraints (e.g., link capacity restrictions). As the network size increases, the number of decision variables and constraints also increases and the computational complexity of the decision-making algorithm becomes the bottleneck.
\begin{table*}[t]
\caption{Different networks considered in this article to which BD is applied.} 
\centering \footnotesize
\begin{tabular}{|p{1.5cm}|p{4cm}|p{3.5cm}|p{4cm}|p{3.5cm}|} 
\hline\hline
\cellcolor{blue!25}&\cellcolor{blue!25} CBD &\cellcolor{blue!25} ComBD &\cellcolor{blue!25} Multi-stage BD for stochastic programs  &\cellcolor{blue!25}GBD\\ 
\hline\hline 
\cellcolor{blue!25} & 				&		&				 & 					 \\
\cellcolor{blue!25} &i. Linear duality				&i. Deals specifically with  integer sub-problems&i. BD applied more than once, recursively & i.  Deals with nonlinear programs specifically  \\	
\cellcolor{blue!25}Key Features & ii. Reformulation with the aid of extreme rays and extreme points &ii. Makes use of combinatorial cuts & ii. Decision of a stage depends on that in the previous stage		 &ii.	Requires sub-problems to be convex \\
\cellcolor{blue!25}                   &												     &							  & iii. Accounts for uncertainties		&iii. Exploits nonlinear duality\\
\cellcolor{blue!25} &										             & 						  &				& iv. Two sub-problems are required for generating the two cut types					\\
\hline
\cellcolor{blue!25} & 				&		&				 & 					 \\
\cellcolor{blue!25}Applications&i. WSNs \cite{WSN1,WSN2}&WLANs\cite{combWLAN}&	 Smart grid with WBS \cite{stochPow} & i. Interworking public WLAN and cellular networks \cite{WLAN1}\\

\cellcolor{blue!25}&ii. Heterogeneous cellular wireless networks\cite{CBD3}&		&	& ii. D2D networks \cite{mypaper, D2D2}\\

\cellcolor{blue!25}&iii. Wired infrastructure WANs \cite{CBD4}&		&						&\\

\cellcolor{blue!25}&					      &	  &				&\\

\cellcolor{blue!25}&					      &	  &					&\\
\hline
\end{tabular}
\label{tax} 
\vspace{0.5cm}
\end{table*}

To design efficient algorithms for large problems, an exploitation of any mathematical structure in the problem's formulation is crucial. An effective approach for that is the use of \textit{decomposition methods}. In this case, the variables and constraints are decomposed into several subsets with the constraints in one subset mainly depending on the variables in the same subset. The idea of decomposition methods is to temporarily fix the linking variables or drop the complicating constraints; hence, the blocks are decoupled and the problem decomposes into several smaller problems.

Benders Decomposition (BD) is one of the popular decomposition schemes \cite{Surv}, because it exploits the structure of the problem and decentralizes the overall computational burden. It has been successfully applied in a variety of network types, as illustrated in Figure \ref{net_types}. Since its introduction \cite{Bender}, several extensions have been proposed for the BD method. In this article, we present the basic concepts of the four main BD variants, as well as some of their various existing applications in communications networks. Moreover, we discuss the technical challenges with BD and possibilities to overcome them when applied to communication networks.


\section{Basics of Benders Decomposition}\label{basic}
The basic idea behind BD is to decompose the problem to be solved into two simpler problems, namely the master problem (MP) and the auxiliary problem (or sub-problem). The MP is a relaxed version of the original problem, containing only a subset of the original variables and the associated constraints. Its solution yields a lower bound on the objective function of the problem. The auxiliary problem is the original problem with the variables obtained in the MP fixed. Its solution yields an upper bound on the objective function of the problem and is used to generate cuts for the MP. The MP and auxiliary problem are solved iteratively, until the upper and lower bound are sufficiently close.

Besides the original BD, also known as \textit{classical Benders decomposition} (CBD), other variants have been developed and successfully applied in different types of data networks problems. These variants include:

\begin{figure}[t!]
	\begin{center}
	\centering
\includegraphics[width=0.99\columnwidth]{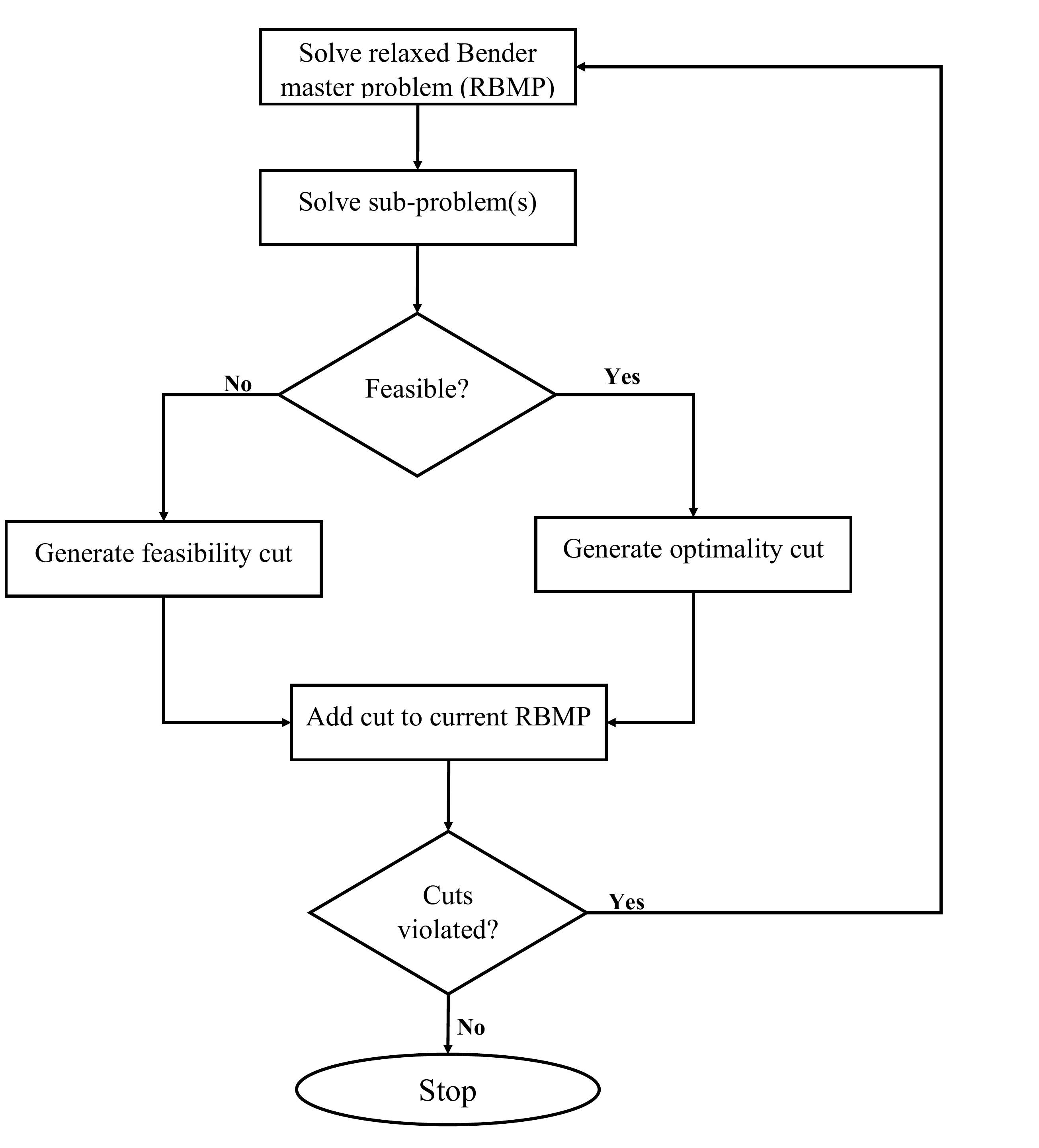}%
\caption{Flowchart representation of BD.}
\label{CBD}
\end{center}
\end{figure}

\begin{enumerate}
	\item Combinatorial BD (ComBD) \cite{comb,combWLAN};
	\item Multistage \textit{l-shaped decomposition} \cite{stoch};
	\item Generalized BD (GBD) \cite{GBD}. 
\end{enumerate}

Figure \ref{CBD} illustrates, at a high level,  an abstract flowchart of how the four
BD variants work. Table \ref{tax} shows different types of communication networks for which a variant of BD has been used to solve a particular design problem, as well as the key features of each variant. 

A given problem can usually be modeled with different equivalent formulations. Formulations have a large impact on the BD performance, as they greatly affect the computational requirements. Formulations with strong \textit{linear program} (LP) relaxations have better computational performance. The reasons are mainly because of the tighter root node, the lower number of fractional variables, and the stronger cuts generated. Tighter formulations can be obtained by adding additional constraints; however, this may result in a more time-consuming MP. Therefore, there is a trade-off between the reduction in the number of iterations and the additional difficulty of the sub-problem.

The BD method has strong ties to other types of decomposition, such as the Dantzig-Wolfe (DW) and Lagrangian optimization, when it comes to solving LPs. Solving an LP by DW is equivalent to solving its dual using BD. The BD is also equivalent to a cutting-plane method applied to the Lagrangian dual of the problem. On the other hand, in integer programming, the relationship between the decomposition methods is not obvious. A great advantage that BD has in this case is that it converges straight to the optimal of the \textit{mixed integer linear program} (MILP) rather than to a relaxation of the problem, as DW decomposition and Lagrangian relaxation do. Therefore, BD needs not be embedded within a branch and bound framework.
\section{Classical Benders Decomposition (CBD)}\label{Sec_CBD}
CBD can be applied for an MILP that has a constraint set connecting both continuous and integer variables linearly. The \textit{Benders master problem} (BMP) is obtained through the following steps:
\begin{enumerate}
\item Reformulating the problem into two nested minimization problems, where the outer minimization is over a discrete variable vector $\mathbf{x_{1}}$, while the inner minimization (IM) is only over a continuous variable vector $\mathbf{x}_{2}$.

\item Interchanging IM with its dual (IMD) using linear duality theory. The feasible region, $F_{IMD}$, of IMD is independent of the choice of $\mathbf{x_{1}}$, which means that if it is non-empty, IMD is either unbounded or feasible for any $\mathbf{x_{1}}$.
\item
\begin{enumerate}[i.]

\item An unbounded direction in $F_{IMD}$ must be avoided, as it indicates an infeasibility of $\mathbf{x}_{1}$.To restrict movement in this direction, a \textit{feasibility} cutting plane must be added. 

\item If the inner IMD is feasible, the solution is an extreme point of $F_{IMD}$.
\end{enumerate}
\item  Another reformulation is done, where an auxiliary variable substitutes the linear IMD objective and a linear constraint set that is function in the vector of the extreme points of $F_{IMD}$ is added. These constraints are the \textit{optimality cuts}, which when added one after the other, tighten the lower bound on the objective function.
\end{enumerate}

Since the complete enumeration of feasibility and optimality cuts is not possible, a relaxation and iterative approach is used to solve the BMP. In a given iteration, the main steps of the iterative procedure are as follows:
\begin{enumerate}
\item The \textit{relaxed} BMP (RBMP) is solved for a subset of cuts to obtain a trial value $\mathbf{x_{1}}=\mathbf{x_{1}^{*}}$.
\item The IMD with $\mathbf{x}_{1}^{*}$ is solved. If the IMD is unbounded, a feasibility cut is generated. Otherwise, an optimality cut is generated.
\item If the cuts are violated by the current solution, they get added to the current RBMP and the process repeats.
\end{enumerate}

CBD has been applied to different existing network design problems. These and their related CBD specifics, are summarized in Table II(a), and discussed next.

\begin{table*}[t!]
\centering
\subfloat[Subtable 1 list of tables text ][Summary of design problems.]{

\includegraphics[width=1.0\textwidth]{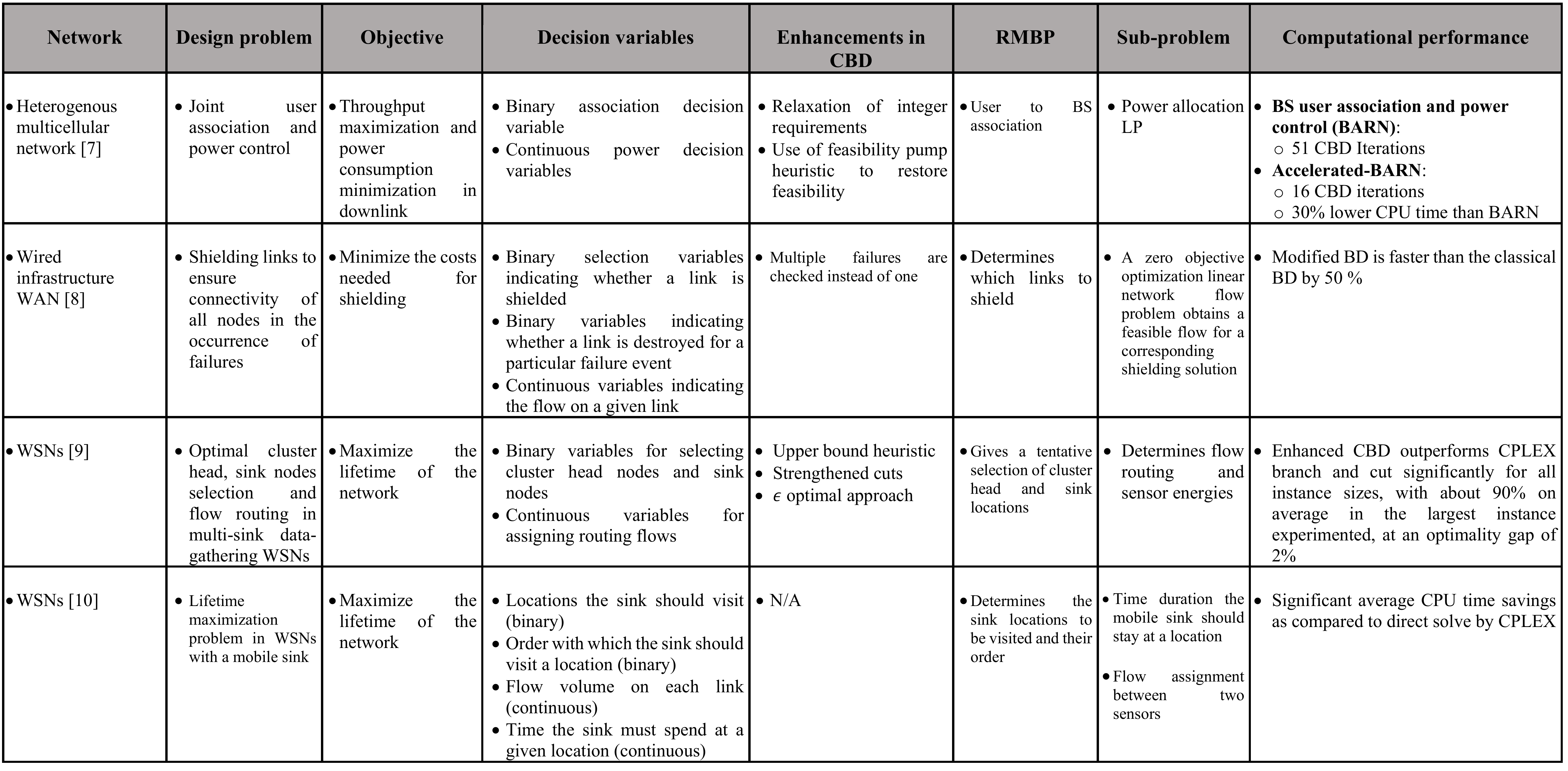}\label{CBD_app}}%
\qquad
\subfloat[Subtable 1 list of tables text ][Running time comparison of CBD and its application-specific enhancement for infrastructure wired WAN link shielding \cite{CBD4}.]{

\begin{tabular}{|c|c|c|} 
\hline\hline
\cellcolor{blue!25} Number of nodes  &\cellcolor{blue!25} CBD: Time (s)  &\cellcolor{blue!25} Application-specific modification: Time (s)\\ 
\hline\hline 
10&2.26&1.54\\
\hline
15&20.89&10.85\\
\hline
20&90.77&34.96\\
\hline
25&270.69&103.32\\
\hline
20&684.83&195.2\\
\hline
\end{tabular}\label{example}}
\caption{Summary of design problems in which CBD is employed.}
\end{table*}

\subsection{CBD for Base Station (BS) Association and Power Control}
CBD is employed for user to BS association and power control in heterogeneous cellular networks in \cite{CBD3}, with the aim of maximizing the overall system revenue measured by either the total data rate or the number of mobile stations in service. The signal-to-interference-and-noise-ratio (SINR) target on each link in service must be satisfied with the minimum total transmission power. It was shown that for certain weighting of the different objectives, the optimal solution of the resulting MILP maximizes the revenue and minimizes the power. The user to BS association is achieved by solving the RBMP iteratively. In the BS association and power control algorithm (BARN) \cite{CBD3}, integer linear programming methods are used for solving the RBMP. In \textit{accelerated} BARN \cite{CBD3}, the integer requirements are relaxed and standard LP solution techniques are used. The feasibility pump heuristic \cite{CBD3} is used to convert the fractional to integer values. The power allocation is done by solving the sub-problem iteratively.
\subsection{CBD for Critical Link Shielding}
A problem that explores the shielding of critical links to enhance the robustness of a network, in which shielded links are resilient to failures, is investigated in \cite{CBD4}. The problem is to find the minimum cost shielding in order to resist both independent random failures and correlated failures, such that connectivity between all source-destination pairs is always maintained.  The problem is modeled as a minimum cost network flow problem that takes into account different failure scenarios. The MP is composed of binary decision variables,  where each indicates whether a link is shielded. The sub-problem is a zero objective optimization linear network flow problem that captures the possible failure patterns and guarantees a feasible flow for a corresponding shielding solution obtained from solving the linear integer RBMP.\\
\textit{Numerical Example}:
As shown/demonstrated in \cite{CBD4}, applying modifications to the CBD can provide significant time enhancement. This is extremely useful because two failures may affect two disjoint sets of links, and the shielding decisions can be made in the same step to reduce the number of iterations. The number of violated constraints added before re-solving the master problem provides a tradeoff between the number of iterations and the running time of each iteration.  Table II(b) shows the corresponding reproduced numerical results where the number of constraints is equal to the number of nodes in the network, which yields 50\% reduction in the running time compared with the standard CBD.

It is hence important to know that there are possibilities to enhance the convergence speed of BD in general, by exploiting application specific features of various communication networks. More on the BD challenges and potential ways to address those is discussed in Section \ref{Fut}.
\subsection{CBD for Data Gathering in WSNs}
Data gathering cluster-based WSNs are studied in \cite{WSN1}, where a WSN with two sets of candidate sink nodes and cluster nodes is considered. The design problem is to find a selection from each set jointly with the network flow between the sensor node and the selected sinks, such that the lifetime of the network is maximized. The objective function represents the weighted sum of three terms: the average energy consumption, the range of remaining energy levels, and the fixed cost associated with locating the cluster heads. The MP is an integer linear program (ILP) which gives a tentative network configuration, i.e., the selection of a cluster head and sink locations. The sub-problem provides the optimal data routing and allocates the sensor energies for that fixed configuration. It is essentially a linear minimization problem that determines the data routing scheme from sensors to sinks via cluster heads and energy usage in the network. Three approaches are used in \cite{WSN1} to speed up the convergence of CBD: an upper bound heuristic algorithm; strengthening the Benders cuts; and following an $\epsilon$-optimal approach.
\subsection{CBD for Routing and Mobile Sink Tour Optimization in WSNs}
In \cite{WSN2}, the problem of maximizing the lifetime of a WSN with a mobile sink is considered. The sink travels at a finite speed among a subset of possible sink locations to collect data from a stationary set of sensor nodes. The problem chooses a subset of sink locations for the sink to visit, finds a tour for the sink among the selected sink locations, and prescribes an optimal data routing scheme from the sensor nodes to each location visited. The sink's tour is constrained by the time it spends at each location collecting data. The CBD starts from a tour over a subset of the sink locations and reaches an optimal solution by adding appropriate Benders cuts. The sink's routing-travel decisions are made by the RBMP, which determines the locations to be visited and their order. The time duration the mobile sink has to stay at a particular location, as well as the flow between two sensors while the sink is at that location, are determined by the sub-problem.

\section{Combinatorial Benders Decomposition (ComBD)}
When a subset of the projected variables must satisfy integrality requirements, the standard linear duality theory is no longer suitable for obtaining classical Benders cuts. A different framework is thus invoked for the cut generation in order to handle integer sub-problems. ComBD is suitable for a basic 0-1 ILP amended by  a set of ``conditional" linear constraints which come into effect if one of the 0-1 variables is `on', i.e., $\mathbf{y}\left(j\right)=1 \Rightarrow \mathbf{a}_{i}^{T}\mathbf{x} \geq b_{i}$ and possibly a set of ``unconditional" constraints, where $\mathbf{x}$ is possibly a vector of continuous variables and $\mathbf{y}$ is a vector of binary variables. For such problems, the MP is a 0-1 integer program and the sub-problem is a feasibility optimization problem with no objective function. ComBD excludes the current MP solution from further consideration via \textit{combinatorial cuts}, which usually take the form
\begin{equation}\label{ComBD_cut}
\sum_{\forall j :\bar{\mathbf{y}}\left(j\right)=0}\mathbf{y}\left(j\right)+\sum_{\forall  j :\bar{\mathbf{y}}\left(j\right)=1}\left( 1-\mathbf{y}\left(j\right) \right )\geq 1,
\end{equation}
and are used as feasiblity cuts in ComBD, where $\bar{\mathbf{y}}$ is the solution returned by the RBMP. 

ComBD was used in \cite{combWLAN} for user association and power control in WLANs. In this problem, each user equipment (UE) must be assigned to one access point (AP). Each AP can have different power levels (PLs) and can be powered off, but at most one PL must be chosen for each AP. The decisions are whether to use or not an AP, to assign a PL to each selected AP, and to assign exactly one AP to each UE. The objective is to minimize the aggregate power of all APs, while satisfying the nonlinear capacity constraints of each AP. The RBMP here solves for the power levels subject to the condition that only one power level per AP is assigned. 

The sub-problem is a zero objective linear integer program. If this is infeasible, a cut as in equation (\ref{ComBD_cut}) can be used as a feasibility cut. The results indicate that ComBD is often orders of magnitude faster than a benchmark cutting planes algorithm.

\section{Multi-Stage Benders Decomposition for Stochastic Programs}
\begin{figure*}[t]
	\begin{center}
	\centering
	\subfloat[Deterministic dynamic problem: There is no uncertainty about the evolution of information. \label{deter}]{
      \includegraphics[width=0.48\textwidth]{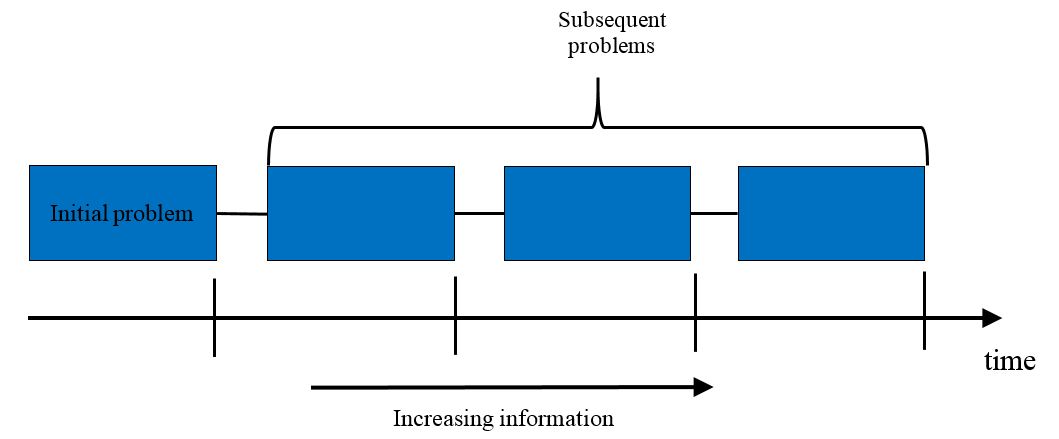}
    }
   \hfill
    \subfloat[Two-stage stochastic dynamic problem. Here there is uncertainty
about which of the three possible successor states the problem moves into after the first decision, such that the decision must consider the uncertainty. \label{stoch}]{
      \includegraphics[width=0.45\textwidth]{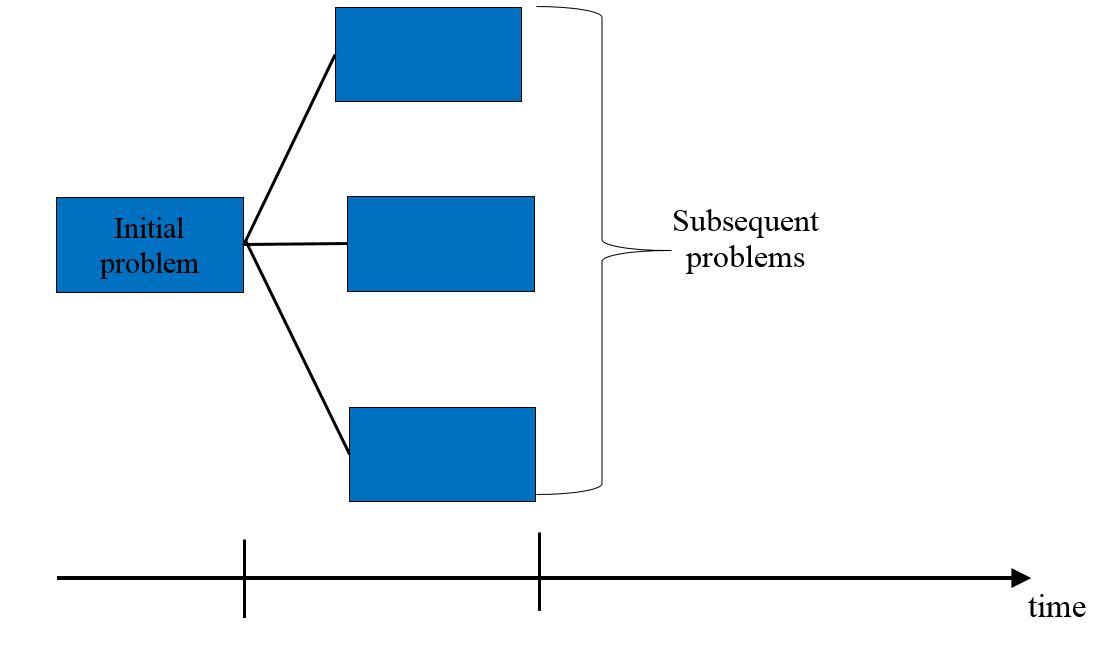}
    }
\caption{Multistage problems.}
\label{multi}
\end{center}
\end{figure*}

%
This technique is suitable for solving \textit{stochastic linear programming} problems, which can be modeled as multi-stage linear programs that consider uncertainty in at least some of the quantities involved in the problem \cite{stoch}. Figure \ref{multi} shows a two-stage recourse problem versus a four stage deterministic problem.

The solution maximizes the expected value of the objective function over all possible scenarios. A multi-stage problem can be formulated as a single optimization problem known as the \textit{deterministic equivalent}. A key approach for that is to discretize the possible future evolution of the state space. This yields a normal linear minimization problem, since the objective function is linear and the constraints are now finite.

\subsection{Applying BD to Stochastic Programming Problems}
The main idea is to split a very large LP problem, which is the deterministic equivalent of the stochastic programming problem, into smaller LP problems that are each much quicker to solve. The decomposition yields a single first stage problem and a number of second stage problems, one corresponding to each of the possible scenarios that could be realized in stage two.

Given an initial estimate of the first-stage variables, an attempt is made to solve each of the sub-problems assuming that this chosen value was the decision made in the first stage. There are two possibilities: either some of the sub-problems are infeasible or all sub-problems are feasible. For the former, feasibility cuts are added to the RBMP, while optimality cuts are added in the latter. 
\subsection{Stochastic Power Management for WBSs in Smart Grids}
An application exists in power management of WBS in a smart grid \cite{stochPow}. The uncertainty is in spot power price of the electrical grid, the power availability in a renewable source, and the demands. The power price from the smart grid is time-varying and increases during the peak hour. The optimization is performed for $\mathcal{\left |T  \right |}$ periods, where $\mathcal{T}$ is a set of time periods and $\left |. \right |$ is the set cardinality. A random variable, $\tilde{\omega}_{t}$,  captures the possible spot power prices from electrical grid, the possible power generated from the renewable power source, and the possible number of connections in the WBS at period $t$. The objective function minimizes the expected cumulative cost over all periods. The period-dependent decision variables are the amount of power bought from an electrical grid, the amount of power stored in a battery, and the amount of excess power to be sold back to an electrical grid at period $t$. The cost at period $t$ is defined with the period-dependent decision variables and composite random variables $\tilde{\omega}_{t}$ and $\tilde{\omega}_{t+1}$. The objective function is a linear function in the decision variables and represents an aggregation of the cost of buying power from the electricity grid, the cost of losing renewable power stored in the battery less the cost of selling power back to the electrical grid. 

The problem formulation renders a stochastic program, which is transformed into a deterministic equivalent formulation that is an LP. The expectation in the stochastic program gets replaced by the weighted sum of probabilities of scenarios. The MP solves for the amount of power to be sold back to the power grid, for certain consumption levels of battery and grid power. The solution obtained from the MP is tested in the sub-problem. If a solution is optimal within the acceptable tolerance, the algorithm stops. Otherwise, the nearly-optimal values of primal and dual variables  from the sub-problems are used to generate Benders cuts that can improve the solution, and another iteration is performed.

\section{Generalized Benders Decomposition (GBD)}\label{GBD}
GBD generalizes CBD to solve nonlinear problems for which the auxiliary sub-problem is a convex program \cite{GBD}. As in CBD, there are two decision vectors $\mathbf{x} \in \mathcal{X}$ and $\mathbf{y} \in \mathcal{Y}$, where $\mathbf{y}$ is a vector of complicating variables in the sense that the problem is a much easier optimization problem in $\mathbf{x}$ when $\mathbf{y}$ is temporarily held fixed. If $\mathbf{y}$ consists of integer decision variables, then the problem is a \textit{mixed integer non-linear program} (MINLP). GBD becomes largely effective if one or more of the following are fulfilled:
\begin{enumerate} 
\item
For a fixed $\mathbf{y}$, the problem divides into a number of independent optimization problems in a separate decision subvector of $\mathbf{x}$;
\item
For a fixed $\mathbf{y}$, the problem assumes a well-known special structure, such as \textit{shortest path tree problem} (STPT) or network flow;
\item
Though the problem is not a convex program in $\mathbf{x}$ and $\mathbf{y}$ jointly, fixing $\mathbf{y}$ renders it so in $\mathbf{x}$.
\end{enumerate}
Nonlinear convex duality is exploited to derive the families of cuts corresponding to those in CBD. In the GBD, to derive the RBMP, projection is invoked to project the problem on to $\mathbf{y}$, leading to $\underset{\mathbf{y}}{min}~\left \{ v\left(\mathbf{y}\right):~~\mathbf{y}\in\mathcal{Y}\cap\mathfrak{V} \right \}$, where $v\left(\mathbf{y}\right)$ is a parametric function that corresponds to the optimum value of the problem under consideration for fixed $\mathbf{y}$, and $\mathfrak{V} \equiv \left\{\mathbf{y}:\mathbf{G}\left(\mathbf{x},\mathbf{y}\right) \leq \mathbf{0},~for~some~\mathbf{x}\in \mathcal{X}\right\}$,
where $\mathbf{G}\left(\mathbf{x},\mathbf{y}\right)\leq \mathbf{0}$ is a constraint set in the original formulation. Non-linear duality is then invoked to derive the BMP, which has two sets of constraints that are both too large and involve an infimum term over $\mathbf{x}\in \mathcal{X}$. The natural strategy for solving the RBMP is ignoring all but a few of its many constraints, then iteratively adding the violated ones to the RBMP.

GBD has been applied in some network design problems. These and their related GBD specifics are summarized in Table \ref{GBD_app}, and discussed in the sequel.
\begin{table*}[t!]
\caption{Summary of design problems in which GBD is employed.}
	\begin{center}
	\centering
\includegraphics[width=0.99\textwidth]{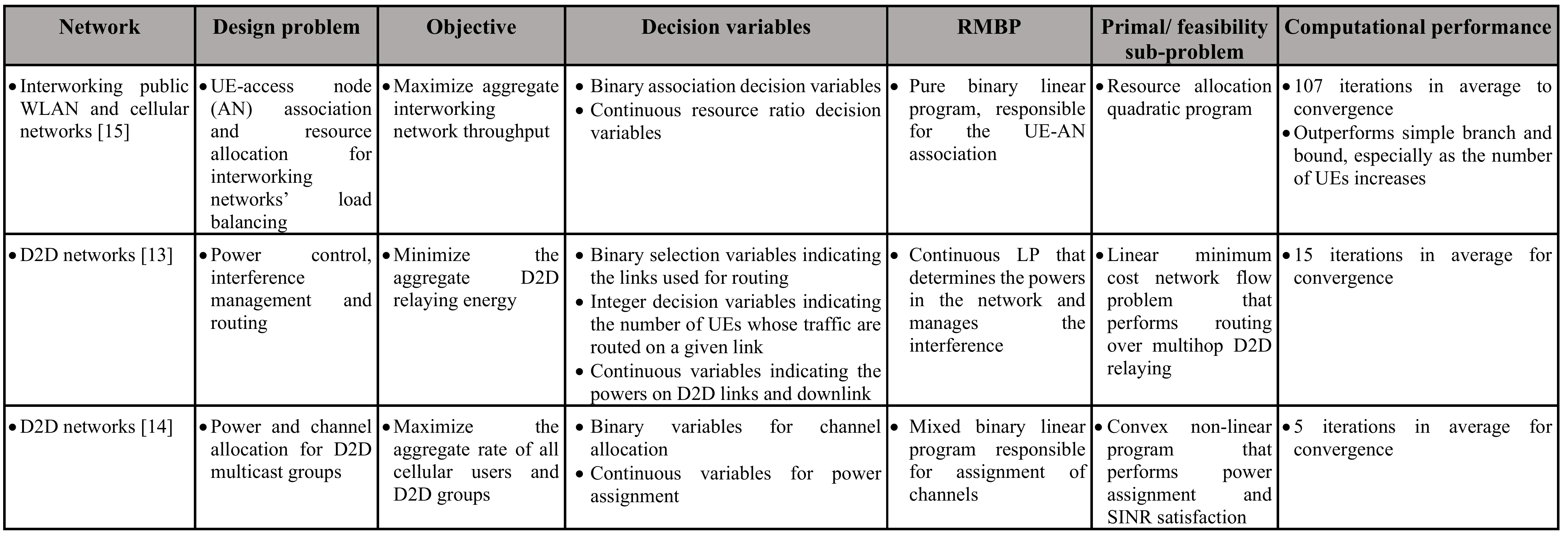}%
\label{GBD_app}
\end{center}
\end{table*}
\subsection{GBD for Joint Power Control and Routing in Multihop D2D Networks}
In our work in \cite{mypaper}, multihop relaying is considered to deliver the packets from the BS to the UEs whose direct link is in outage, i.e., whose signal-to-noise-ratio (SNR) is below an acceptable threshold. There is interference from the D2D relay transmissions on uplink cellular users, which must be managed. Furthermore, the resource blocks (RBs) for D2D sidelinks (SLs) are reused across the cell, so the SL interference must also be controlled. Virtual circuit routing is considered from BS to UEs in the cell. The objective is to minimize the aggregate SL energies expended on all RBs while routing the transmissions from the BS to all UEs in the network. Downlink SNR and SL SINR thresholds must be satisfied for all RBs. The problem is an MINLP with a bilinear mixed integer objective function.

The RBMP is an LP in the power decision variables that grows by one additional constraint in each GBD iteration. Dual simplex and sensitivity analysis are used to exploit this by using the RBMP's solution from the previous iteration to reduce the computational effort resulting from its growing size.  After necessary equivalent reformulations, the auxiliary sub-problem turned out to be an LP. When partial Lagrangian is applied to the sub-problem, it results in a linear minimum cost network flow program (MCNFP) structure and another trivial mixed binary linear problem with box constraints only. The network simplex is used to solve the resulting MCNFP. In case the primal sub-problem is infeasible, an $l_{1}$-norm feasibility sub-problem is solved. 
\subsection{GBD for Load Optimization in Cellular and Public WLAN Interworking Network}
In \cite{WLAN1}, public WLAN is exploited to expand the capacity of existing cellular access networks. A main problem that is addressed is the underutlization of the public WLAN APs. For the amelioration of the traffic load, a centralized optimization model that jointly optimizes UE-access node (AN) association and resource allocation to maximize the overall throughput with some throughput fairness considerations, is proposed. Users can access only one AN simultaneously, either a cellular BS or a WLAN AP, depending on the strength of the received power. The association is represented by binary decision variables, while the resource share of the cellular network BS is modeled by a continuous variable. The objective function is mixed integer quadratic in the decision variables of the UE-AN association and the BS resource ratio share. 

The problem is mixed integer quadratic program. The RBMP is a pure binary linear program, which is responsible for the UE-AN association. In each iteration, a tentative association is tested for feasibility in a continuous non-linear quadratic sub-problem, which generates an optimality cut, if feasible. If this is not feasible, a cut of the type in (\ref{ComBD_cut}) is used as a feasibility cut.
\subsection{GBD for Joint Power Control and Channel Allocation in Multicast D2D Networks}
Multicast D2D communications underlaying a cellular network is investigated in \cite{D2D2}. In this problem, each D2D group can reuse the uplink channels of multiple cellular users, and the channel of each cellular user can be reused by multiple D2D groups. A variant of GBD is used for solving an MINLP to find a joint power control and channel allocation solution that maximizes the aggregate rate of all cellular users and D2D groups. Minimum SINR requirements are imposed by constraints. The transmission powers are continuous variables and the channel allocations are modeled by binary variables. The MINLP is decomposed into a sub-problem and an RBMP: the former is a convex NLP that corresponds to the original problem with fixed binary variables, and the latter is a mixed binary LP derived through nonlinear duality theory using Lagrange multipliers obtained from the former. The auxiliary sub-problem is responsible for the power allocation and satisfaction of SINR requirements, while the RBMP is responsible for the subchannel assignments. In case the auxiliary sub-problem is infeasible, an $l_{1}$-norm feasibility sub-problem is solved.


\section{Challenges and Future Directions}\label{Fut}
The aforementioned variants of BD provide vast capabilities in developing a tractable solution methodology for different design problems in various communication networks. Challenges in BD include \cite{Surv}: time-consuming iterations; poor feasibility and optimality cuts; low effectiveness of first few iterations; zigzagging primal solutions; slow convergence at a large number of iterations near the optimal solutions; upper bounds that do not improve for many iterations because multiple solutions with the same objective function value exist.

Improving the convergence can be obtained by reducing the number of iterations and the time required per iteration \cite{Surv}. The former can be achieved by improving both the quality of the generated solutions and the cuts. The later can be obtained by exploiting structures in the MP and the auxiliary sub-problem. In what follows, we briefly discuss some of the possible enhancement strategies for BD.

\subsection{Decomposition Strategy}
This determines how the problem is projected to get the initial MP and the auxiliary sub-problem(s). All the linking constraints and non-complicating variables are projected out in the \textit{conventional} decomposition. A discussion was given in \cite{partial_project} on how this causes the MP to lose all the information associated with non-complicating variables. As a result, there can be instability, a large number of iterations, and irregular progression of bounds. \textit{Partial BD} strategies were proposed in \cite{partial_project} that add to the master explicit information from the sub-problems, either by retaining or creating scenarios, or even both. Significant improvements in terms of computational time and number of generated cuts were obtained. Researchers designing decision-making algorithms in communication networks can explore ways to exploit this, especially if the problem is a stochastic program.

\subsection{Solution Procedure}
This is specifically related to the algorithms used for the MP and the auxiliary sub-problem(s). The common standard techniques would be the simplex and branch-and-bound algorithms, which are usually treated as blackbox solvers. Advanced techniques make use of the structure of the MP and the auxiliary sub-problem or the search requirements of the BD algorithm. It was reported that 90\% of the BD computational effort is needed for solving the MP. 

Size management of the MP can be done by removing unnecessary cuts or avoiding the insertion of available cuts. Efficient solution methods for the MP include:
\begin{itemize}
\item
Solving the MP to $\epsilon$-optimality, especially at the early iterations of the BD.
\item
Solving the MP using meta-heuristics.
\item
Using \textit{constraint programming}  to handle special constraints more efficiently.
\item Using \textit{column generation} to handle certain structures effectively. 
\end{itemize}
For the sub-problem, specialized algorithms can be used (e.g., transportation simplex, Hungarian algorithm, second-order cone programming) if it has a special structure.
\subsection{Solution Generation}
This has to do with the method for setting the trial values of the complicating variables. \textit{Heuristics}, \textit{alternative MP} or an \textit{improved MP} can be exploited to generate solutions more quickly or for obtaining better solutions. \textit{Hybrid} approaches can also be used. Even simple heuristics can generate high-quality initial solutions and cuts, repair infeasible solutions and reduce the computational cost of the MP and sub-problem. Finally, altering the MP, even temporarily, can mitigate slow generation of solutions (possibly with low qualities) and instability. 
\subsection{Cut Generation}
This is mainly related to the technique used for the generation of optimal and feasibility cuts. The conventional approach does so by solving the auxiliary (or feasibility) sub-problems obtained from the decomposition. This can be inefficient for solving certain problems where the sub-problems are degenerate. Therefore, other possible strategies would invoke reformulations of the sub-problems.

\section{Conclusion}\label{Conc}
This article reviewed the main variants of BD with existing applications to various communications networks. The BD basics were introduced and the impact of formulation on the computational performance was discussed, along with the ties of BD to other existing decomposition schemes. Each BD variant was presented along with with some of its existing applications including D2D networks, WSNs, WLANs, and infrastructure wired networks. Existing challenges in BD and ways in which these can be addressed were also discussed. In a nutshell, BD is a powerful technique that can find numerous applications to various network design problems.

\bibliographystyle{ieeetr}

\vspace{-1cm}
\begin{IEEEbiographynophoto} 
{Ahmed Ibrahim} (M’19) is currently a post-doctoral fellow and a per course instructor with the Memorial University of Newfoundland. He received his Ph.D. degree in electrical and computer engineering from the University of Manitoba (UoM) in October 2016. He coauthored a book entitled Optimization Methods for User Admissions and Radio Resource Allocation for Multicasting over High Altitude Platforms published by River Publishers in February 2019. He was a recipient of the Best Paper Award from the IEEE Wireless Communications and Networking Conference (WCNC) in 2019. He was awarded as an Exemplary Reviewer of the IEEE COMMUNICATIONS LETTERS journal in 2017.
\end{IEEEbiographynophoto}
\vspace{-1cm}
\begin{IEEEbiographynophoto} 
{Octavia A Dobre} (M’05–SM’07) is a Professor and Research Chair at Memorial University, Canada. She was a Visiting Professor at Massachusetts Institute of Technology, as well as a Royal Society and a Fulbright Scholar. Her research interests include technologies for 5G and beyond, as well as optical and underwater communications. She published over 250 referred papers in these areas. Dr. Dobre serves as the Editor-in-Chief (EiC) of the IEEE Open Journal of the Communications Society. She was the EiC of the IEEE Communications Letters, a senior editor and an editor with prestigious journals, as well as General Chair and Technical Co-Chair of flagship conferences in her area of expertise. She is a Distinguished Lecturer of the IEEE Communications Society and a fellow of the Engineering Institute of Canada.
\end{IEEEbiographynophoto}
\vspace{-1cm}
\begin{IEEEbiographynophoto}
{Telex M. N. Ngatched }  (M’05–SM’17) is an Associate Professor and Coordinator of the Engineering One program at the Grenfell Campus at Memorial University, Canada. His research interest include 5G enabling technologies, cognitive radio networks, visible light and power-line and communications, optical communications for OTN, and underwater communications. Dr. Ngatched is an Associate Editor of the IEEE Open Journal of the Communications Society and the Managing Editor of IEEE Communications Letters. He was the publication chair of IEEE CWIT 2015, an Associate Editor with the IEEE Communications Letters from 2015 to 2019, and Technical Program Committee (TPC) member and session chair for many prominent IEEE conferences in his area of expertise.
\end{IEEEbiographynophoto}

\begin{IEEEbiographynophoto}
{Ana Garcia Armada} (S’96–A’98–M’00–SM’08) is a Professor at University Carlos III of Madrid, Spain. She has published around 150 referred papers and she holds four patents. She serves on the editorial board of IEEE Communications Letters, IEEE Trans. on Communications and IEEE Open Journal of the Communications Society. She has served on the TPC of more than 50 conferences and she has been part of many organizing committees. She has received several awards from University Carlos III of Madrid, including an excellent young researcher award and an award to best practices in teaching. She was awarded the third place Bell Labs Prize 2014 for shaping the future of information and communications technology. Her research mainly focuses on signal processing applied to wireless communications.
\end{IEEEbiographynophoto}
\end{document}